\documentclass{ws-procs9x6}

\def\beq{\begin{eqnarray}}
\def\eeq{\end{eqnarray}}
\def\bea{\begin{eqnarray}}
\def\eea{\end{eqnarray}}
\newcommand{\nn}{\nonumber}
\newcommand{\nats}{\mbox{${\rm I\!N }$}}

\usepackage{color}

\begin{document}

\title{Semitransparent pistons}

\author{P. Morales and K. Kirsten}

\address{Department of Mathematics, Baylor University,\\ Waco, Texas 76798-7328, USA\\
E-mail: pedro\_morales@baylor.edu and klaus\_kirsten@baylor.edu}

\begin{abstract}
We consider semitransparent pistons in the presence of extra dimensions. It is shown that the piston is always attracted to the closest wall irrespective of details of the geometry and topology of the extra dimensions and of the cross section of the piston. Furthermore, we evaluate the zeta regularized determinant for this configuration.
\end{abstract}


\bodymatter

\section{Introduction}\label{sec1}
In this contribution we consider three-dimensional pistons of arbitrary cross section in the context of Kaluza-Klein models. As is well known, pistons have the important advantage that they allow for an unambiguous prediction of Casimir forces \cite{cava04-69-065015}. This is the main reason for the recent surge of interest in these configurations; see, e.g., refs. [3-7]\nocite{hert05-95-250402,mara07-75-085019,teo09-819-431,kirs09-79-065019,eliz09-79-065023}. Most of the research done so far has concentrated on a rectangular cross section with boundary conditions that allow for an explicit determination of the energy eigenvalues for the configuration, at least for part of the spectrum. Here we want to investigate further geometries with arbitrary cross section, along the lines of refs. [5,7]\nocite{teo09-819-431,kirs09-79-065019}, with boundary conditions leading to a transcendental equation for the spectrum \cite{eliz09-79-065023}. Specifically, we will consider semitransparent pistons and we will show that the piston is attracted to the closest wall. This statement holds independently of the cross section of the piston and of the geometry and topology of the additional Kaluza-Klein dimensions.
\section{Zeta function for semitransparent pistons}
Let $M=[0,L]\times N$, where $N$ represents the cross section of the piston and the additional Kaluza-Klein dimensions assumed to be a smooth Riemannian manifold possibly with a boundary. We place Dirichlet plates at $x=0$ and $x=L$ and the semitransparent piston at $x=a$ is modeled by a delta function potential.
The energy eigenvalues for a scalar field are then determined by the second order differential operator
\beq P = - \frac{\partial^2}{\partial x^2} - \Delta_N + \sigma \delta (x-a),\label{1}\eeq
together with Dirichlet boundary conditions at $x=0$ and $x=L$. The operator $\Delta_N$ can be thought of as being the Laplacian on $N$, but it might also contain a term showing a coupling to the curvature on $N$ as well as a mass term.

Using separation of variables, eigenfunctions, namely solutions of the equation $$P \phi (x,y) = \lambda^2 \phi (x,y),$$ are written in the form $$\phi (x,y) = X (x) \varphi (y), \quad \quad x\in [0,L], \quad y\in N. $$ Assuming $\varphi (y)$ to be an eigenfunction of $-\Delta _N$, that is $$ -\Delta _N \varphi _\ell  (y) = \eta_\ell ^2 \varphi_\ell (y) , $$ with boundary conditions imposed if $\partial N \neq \emptyset$, the Dirichlet condition along the $x$-axis implies $X(0) = X(L) =0$. In addition, we impose continuity at $x=a$, namely $X (a+) = X(a-)$, and the presence of the delta function $\delta (x-a)$ creates a jump in the derivative, $X' (a+) - X' (a-) = \sigma X (a)$. Incorporating all the above information, eigenvalues $\lambda^2$ are seen to be of the form $\lambda_{k\ell} = \nu_k^2 + \eta _\ell ^2$, where the $\nu_k$ satisfy the transcendental equation \beq\sigma \sin (\nu a) \sin (\nu [L-a]) + \nu \sin (\nu L) =0.\label{2}\eeq The advantage of this particular representation of the secular equation is that in the limit $\sigma \to 0$ we immediately obtain the answer for the configuration of two parallel plates at distance $L$.

Applying the contour integral formulation of zeta functions put forward in ref. [1]\nocite{bord96-37-895}, the zeta function reads $$\zeta (s) = \sum_{k,\ell} (\nu_k^2+\eta _\ell ^2) ^{-s} = \frac 1 {2\pi i} \sum_\ell \int\limits_\gamma d\nu (\nu^2+\eta_\ell^2)^{-s} \frac d {d\nu} \ln F(\nu ), $$ where $\gamma$ is a contour enclosing all positive solutions of eq. (\ref{2}) and \beq F(\nu ) = \frac 1 {\nu^2} \left( \sigma \sin (\nu a) \sin (\nu [L-a]) + \nu \sin (\nu L) \right). \label{3}\eeq In writing $F(\nu )$, the transcendental eq. (\ref{2}) has been divided by $\nu^2$ in order to avoid contributions from the origin in the contour manipulations to come. After deforming the contour to the imaginary axis the representation reads \beq \zeta (s) = \frac{\sin \pi s} \pi \sum_\ell \int\limits_{\eta_\ell}^\infty dk (k^2-\eta_\ell^2)^{-s} \frac d {dk} \ln F(ik) \label{4}\eeq and our next task is to construct the analytical continuation to a half-plane containing the points $s=-1/2$ (for the Casimir force) and $s=0$ (for the functional determinant). As usual, the asymptotic $k\to \infty$ behavior of $F(ik)$ plays the dominant role. Furthermore, the zeta function $\zeta_N (s)$ related to the transversal dimensions,
 \beq \zeta_N (s) = \sum_\ell \eta_\ell ^{-2s}, \nn\eeq will make its appearance. For notational simplicity we assume $\eta_\ell^2>0$.

We first note that $$F(ik) = \frac 1 {k^2} \left( \frac \sigma 4 + \frac k 2 \right) e^{kL} \left[ 1+E_1 (k) \right], $$ where $E_1 (k)$ is exponentially damped as $k\to\infty$. This shows, as $k\to\infty$, $$\ln F(ik) = kL - \ln (2k) + \sum_{j=1}^\infty (-1)^{j+1} \left( \frac \sigma {2k} \right)^j \frac 1 j + E_2 (k) ,$$ $E_2 (k)$ denoting exponentially damped terms. Subtracting and adding the leading $M+2$ terms in this expansion, the zeta function $\zeta (s)$ is naturally split into two pieces,
$$\zeta (s) = \zeta _f (s) + \zeta _{as} (s), $$ where \beq \zeta _f (s) &=& \frac{\sin \pi s} \pi \sum_\ell \int\limits_{\eta _\ell}^\infty dk (k^2 - \eta _\ell ^2) ^{-s} \times \nn\\
& &\hspace{0.5cm}\frac d {dk} \left[ \ln F(ik) - kL + \ln (2k) - \sum_{j=1}^M (-1)^{j+1} \left( \frac \sigma {2k} \right)^j \frac 1 j\right], \label{5}\eeq
and after performing the $k$-integration $\zeta _{as} (s)$ reads,
\beq \zeta_{as} (s) &=& \frac{L\Gamma \left( s-\frac 1 2 \right)}{2 \sqrt \pi \Gamma (s)} \zeta_N \left( s- \frac 1 2\right) - \frac 1 2 \zeta_N (s) \nn\\ & & + \sum_{j=1}^M (-1)^j \left( \frac \sigma 2 \right)^j \frac{\Gamma \left( \frac j 2 +s\right)}{\Gamma \left( 1+ \frac j 2 \right) \Gamma (s)} \zeta_N \left( s+ \frac j 2 \right).\label{6}\eeq
As follows from the $k\to\eta_\ell$ and $k\to\infty$ behavior, the representation for $\zeta_f (s)$ is valid for $1>\Re s > (n-M-1)/2$ where $n=\mbox{dim} (N)$. In the following, choosing $M=n$ respectively $M=n+1$ we will obtain the results for the functional determinant respectively the Casimir force for the situation under consideration.

Let us first evaluate $\zeta ' (0)$ and thus we put $M=n$. The result for $\zeta_f ' (0)$ is trivially obtained as the integral occurring in (\ref{5}) is analytic about $s=0$. We find \beq \zeta _f ' (0) = \sum_\ell \left( \ln F(i\eta_\ell ) - \eta _\ell L + \ln (2\eta _\ell ) + \sum_{j=1}^n (-1)^j \left( \frac \sigma {2\eta _\ell } \right)^j \frac 1 j \right) .\label{7} \eeq In order to explicitly evaluate this expression, once the manifold $N$ is specified, the eigenvalues $\eta_\ell$, if known explicitly, would be substituted and (\ref{7}) evaluated numerically. If the eigenvalues $\eta_\ell$ can only be determined numerically, then a suitably large number of $\eta_\ell$'s needs to be determined numerically and again (\ref{7}) needs to be evaluated numerically.

For the evaluation of $\zeta_{as} ' (0)$ we note that the zeta function $\zeta_N (s)$ has poles at $s=-(2j+1)/2,$ $j\in\nats$, and at $s=1/2,1,...,n/2$. As is well known, about singular points $s=k$ we have the expansion $$\zeta _N \left( s+ k  \right) = \frac 1 s \mbox{Res } \zeta_N ( k)
+ \mbox{FP } \zeta_N (k)+{\cal O} (s).$$ As a result we find \beq \zeta_{as} ' (0) &=& - L \left( \mbox{FP } \zeta_N \left( - \frac 1 2 \right) - \mbox{Res } \zeta_N \left( - \frac 1 2 \right) [ -2+\ln 4] \right) - \frac 1 2 \zeta_N ' (0) \nn\\
& & \hspace{-1.5cm}+ 2 \sum_{j=1}^n (-1)^j \left( \frac \sigma 2 \right)^j \frac 1 j \left\{ \mbox{FP } \zeta_N \left( \frac j 2 \right) + \mbox{Res } \zeta_N \left( \frac j 2 \right) \left[ \gamma + \psi \left( \frac j 2 \right) \right] \right\} ,\label{8} \eeq
with the Euler-Mascheroni constant $\gamma$ and the psi function $\psi (x) = \Gamma' (x) / \Gamma (x)$.
This is as far as we can go without specifying the manifold $N$. Once $N$ is specified, for example as a torus or sphere, the quantities appearing in (\ref{8}) can be evaluated explicitly.

The evaluation of the Casimir force $$F_{Cas} = - \frac 1 2 \frac \partial {\partial a} \zeta \left( - \frac 1 2\right)$$ is simplified by the observation that the asymptotic terms do not depend on the distance $a$. As a result only $\zeta _f (s)$ contributes to $F_{Cas}$ and a representation for the force is \beq F_{Cas} = \frac 1 {2\pi} \sum_\ell \int\limits_{\eta_\ell}^\infty dk (k^2-\eta_\ell ^2) ^{1/2} \frac \partial {\partial a} \frac \partial {\partial k } \ln F(ik) .\nn\eeq
More explicitly we have that \beq h(k) &=& \frac \partial {\partial a} \ln F(ik) = \frac{\sigma k \sinh (k [L-2a])}{\sigma \sinh (ka) \sinh (k [L-a])+k \sinh (kL)} \nn\\
&=& \left( \frac {\sinh (kL)} {\sigma \sinh (k [L-2a])} + \frac {\sinh (ka) \sinh (k [L-a])}{k \sinh (k [L-2a])} \right)^{-1} .\label{9}\eeq
Noting that $g(k) = \sinh (ka) /k$ as well as $f(k) = \sinh (mk) / \sinh (nk)$ for $m>n>0$ is increasing for $k>0$, we conclude that for $0<a<L/2$ the function $h(k)$ is a decreasing function of $k$, whereas for $L/2<a<L$ it is an increasing function of $k$. This shows the piston is always attracted to the closest wall as was found for Dirichlet boundary conditions on the piston and the plates at $x=0$ and $x=L$\cite{kirs09-79-065019}.\\[-.1cm]

{\bf Acknowledgement:} The authors would like to thank Kimball Milton and Steve Fulling for very helpful suggestions.
KK is supported by National Science Foundation grant PHY--0554849.


\end{document}